\documentclass[reprint, aps, pre,  superscriptaddress]{revtex4-1}

\usepackage{amsmath}
\usepackage{graphicx}
\usepackage{epstopdf}
\usepackage{color}
\usepackage{amssymb}
\usepackage{amsmath}
\usepackage{tabularx}
\usepackage{bm}
\usepackage{hyperref}
\usepackage{ltxtable}
\usepackage{natbib}
\usepackage{longtable}
\usepackage[normalem]{ulem}
\usepackage{float}
\usepackage{graphicx}
\usepackage{feynmf}
\usepackage{tikz}
\usepackage[compat=1.1.0]{tikz-feynman}

\newcommand{\la}{\left\langle}
\newcommand{\ra}{\right\rangle}
\newcommand{\be}{\begin{equation}}
\newcommand{\ee}{\end{equation}}
\newcommand{\bse}{\begin{subequations}}
\newcommand{\ese}{\end{subequations}}
\newcommand{\bea}{\begin{eqnarray}}
\newcommand{\eea}{\end{eqnarray}}
\newcommand{\ba}{\begin{array}}
\newcommand{\ea}{\end{array}}

\begin{document}

\title{Renormalization of shell model of turbulence}
\author{Mahendra K. Verma}
\email{mkv@iitk.ac.in}
\affiliation{Department of Physics, Indian Institute of Technology Kanpur, Kanpur 208016, India}
\author{Shadab Alam}
\email{shadab@iitk.ac.in}
\affiliation{Department of Mechanical Engineering, Indian Institute of Technology Kanpur, Kanpur 208016, India}

\date{\today}

% keywords: pattern formation, coarsening, energy transfer
\begin{abstract}
Renormalization enables a systematic scale-by-scale analysis of multiscale systems. In this paper, we employ \textit{renormalization group} (RG) to the shell model of turbulence  and show that the RG equation is satisfied by $ |u_n|^2 =K_\mathrm{Ko} \epsilon^{2/3} k_n^{-2/3}$ and $ \nu_n  =  \nu_* \sqrt{K_\mathrm{Ko}} \epsilon^{1/3} k_n^{-4/3}$, where $k_n, u_n $ are the wavenumber and velocity of shell $ n $; $\nu_*, K_\mathrm{Ko}$ are RG and Kolmogorov's constants; and $ \epsilon $ is the energy dissipation rate. We find that $\nu_* \approx 0.5$ and  $K_\mathrm{Ko} \approx 1.7$, consistent with earlier RG works on Navier-Stokes equation. We  verify the theoretical predictions using numerical simulations. 
\end{abstract}
\maketitle

\section{Introduction}

%Earliest RG calculations for the charge and mass renormalization in quantum electrodynamics (QED) were perturbative~\cite{Peskin:book:QFT,Lancaster:book,Zinn-Justin:book}. Later \citet{Wilson:PRD1971,Wilson:RMP1983} formulated a powerful and intuitive RG scheme, which is referred to as Wilson \textit{exact renormalization group equation} (ERGE). This scheme involves a sharp wavenumber cutoff. \citet{Polchinski:NPB1984}  proposed a new ERG method where he employed a smooth ultraviolet cutoff and reorganized the effective Lagrangian into relevant and irrelevant parts.  Since then, exact renormalization has been performed for many physical systems, including  $\phi^4$ theory~\cite{Polchinski:NPB1984,Berges:PR2002}, chaos~\cite{Hu:PRL1982}, turbulence~\cite{Tomassini:PLB1997,Fontaine:arxiv2022,Canet:JFM2022}, QED~\cite{Gies:PRL2004}, etc.  Note  that in exact renormalization, the  flow equation is reasonably straightforward to write, however, the computation of renormalized parameters is quite involved (see e.g.,~\cite{Tomassini:PLB1997,Fontaine:arxiv2022,Canet:JFM2022}). We also remark that ERG has a similar framework as nonperturbative RG and functional RG. Refer to \cite{Wilson:PR1974,Berges:PR2002,Bervillier:CMP2013,Rosten:PR2012,Calzetta:book,Berges:CP2004,Delamotte:book_chapter} for more details on these topics.

Renormalization group (RG) analysis has been employed to model turbulence. \citet{Orszag:CP1973} and \citet{Forster:PRA1977} performed one of the first perturbative renormalization analysis. \citet{Yakhot:JSC1986} performed detailed analysis using $ \epsilon $ expansion. The other perturbative RG works are by  \citet{Zhou:PRA1989,Zhou:PR2010}, \citet{McComb:PRA1983,McComb:JPA1985,McComb:book:RG,McComb:book:HIT}, \citet{Eyink:PRE1993}, \citet{Martin:PRA1973}, \citet{Bhattacharjee:PF1991}, and  \citet{Adzhemyan:book:RG}.  Among these works,  McComb, Zhou, and coworkers employed self-consistent RG (using ``dressed Green's function") that has   nonperturbative features.  The above set of works show  that the renormalized turbulent  viscosity   $ \nu(k) \sim k^{-4/3} $, where $ k $ is the wavenumber. 

Recently, researchers have employed \textit{exact renormalization group equation} (ERGE) to turbulence~\cite{Wilson:PRD1971,Wilson:RMP1983,Polchinski:NPB1984}. Here, either sharp or smooth filter is employed during coarsening.  A more formal implementation of ERGE is via functional renormalization group (FRG). 
 \citet{Tomassini:PLB1997}, \citet{Fontaine:arxiv2022}, and \citet{Canet:JFM2022}  employed FRG to hydrodynamic turbulence and shell model. They derived formulas for the velocity correlations and multiscaling exponents. For Navier-Stokes equation, \citet{Canet:JFM2022}  reported $ \nu(k) \approx k^{-1} $, rather than $ \nu(k) \sim k^{-4/3} $.  {   \citet{Fedorenko:JSM2013} performed FRG to decaying Burgers, hydrodynamic, and quasi-geostrophic turbulence. Among many results, \citet{Fedorenko:JSM2013} showed that for hydrodynamic turbulence, the second-order structure function scales as $l$ (the distance between two points), rather than Kolmogorov's predictions of $l^{2/3}$. 
 
\citet{Mejia:PRE2012} performed nonperturbative renormalization group analysis of stochastic Navier-Stokes equation with
power-law forcing. Here, they renormalized the viscosity, the forcing amplitude, and the coupling constants. Using field-theoretic tools, \citet{Biferale:PRE2017} constructed optimal subgrid closure for the shell models; they related the closure scheme  to \textit{ large-eddy simulations}.}  In addition, \citet{Eyink:PRE1993} used \emph{operator product expansion} (OPE) and discovered \textit{multiscaling} for the shell model.  Some other notable field-theoretic works (not RG) on turbulence are \cite{Kraichnan:JFM1959,Leslie:book,Fairhall:PRL1997,Falkovich:RMP2001}.

{  In this paper, we employ RG scheme based on the differential equation, as in \cite{Yakhot:JSC1986,McComb:book:RG,Zhou:PRA1989}.     Note that the shell model involves discrete wavenumbers, hence its renormalization does not involve complex integration, as in hydrodynamic turbulence. For inviscid shell model, our RG procedure yields $ \nu(k) = 0 $ as the solution of the RG equation, which is similar to the Gaussian fixed point of Wilson $ \phi^4 $ theory~\cite{Wilson:PR1974}.  We verify several RG predictions using numerical simulation of the shell model.   We use temporal autocorrelation function for the velocity field to compute the renormalized viscosity~\cite{Sanada:PF1992,Verma:INAE2020_sweeping}. 

 In one of the important works on hydrodynamic turbulence,  \citet{Kraichnan:PF1964Eulerian} argued that large-scale structures sweep the small-scale fluctuations; this phenomenon is referred to as \textit{sweeping effect}. These interactions are naturally multiscale (across many wavenumbers). Note, however, that multiscale interactions are absent in the shell models, which has local interactions among the  wavenumber shells. Hence, we expect that sweeping effect may be suppressed in the shell model. This is precisely what we observe in our RG calculation of the shell model. }
 
In this paper, we compute  the renormalized viscosity in the shell model using  momentum-space RG proposed by Wilson~\cite{Wilson:PR1974}  (see Sec.~II).    Here, we assume that the coarse-grained velocity field is random  satisfying  time-stationarity. Our calculation does not require quasi-Gaussian approximation for the velocity field.  In Sec.~III, we compute the energy flux of the shell model; here, we assume the velocity field to be quasi-Gaussian.  The flux calculation enables us to compute  Kolmogorov's constant.   Interestingly, our predictions for the shell model are quite close to those for the Navier-Stokes equation. In Sec.~IV, we extend our RG calculation to show that sweeping effect is suppressed in the shell model.

In Sec.~V, we describe how we verify the theoretical predictions using numerical simulations. We observe that the numerical results are  in good agreement with the theoretical predictions. In Sec.~VI, we compare our results with those from earlier works. We conclude in Sec.~VII.

\section{Renormalization of viscosity}

{  The Sabra shell model is~\cite{Lvov:PRE1998, 	Lvov:EPL1999,Biferale:ARFM2003,Constantin:PD2006,Ditlevsen:book,Plunian:PR2012} 
\bea
\frac{d u_n}{dt} + \bar{\nu} k_n^2 u_n  & = & -i \lambda [a_1 k_n u_{n+1}^* u_{n+2} +  a_2 k_{n-1} u_{n-1}^* u_{n+1}
 		\nonumber \\
 && ~~~ - a_3 k_{n-2} u_{n-1} u_{n-2}] + f_n,
 \label{eq:shell_model}
\eea
where $u_n$ represents the velocity field for the shell $n$; $ \bar{\nu} $ is the \textit{microscopic kinematic  viscosity}; $ a_1, a_2, a_3$ are constants with $ a_1 + a_2 + a_3 =0 $; and $ k_n = k_0 b^n $ with $ b $ as a constant.   In this paper, we choose $ a_1 = 1, a_2 = -1+1/b $,  $ a_3 = -1/b $, and $ b $ in the range $ (1.2,2) $. Here, $f_n$ represents the forcing, which is employed at small $n$'s. This forcing injects energy at large scales that cascades to small scales as the energy flux.  Note that triadic interactions of hydrodynamic turbulence are modelled better  with Sabra model than GOY model~\cite{Lvov:PRE1998}. }

The coupling constant (coefficient of the nonlinear term) $ \lambda $ is not renormalized due to the Galilean invariance~\cite{Forster:PRA1977,McComb:book:Turbulence,McComb:book:HIT}, and we set $ \lambda = 1 $.  Refer to Appendix A for details.  In addition, we consider  $ u_n $ to be random, as in fully-developed turbulence, rather than introducing a separate noise term in the inertial range~\cite{McComb:PRA1983,Zhou:PRA1988,McComb:book:Turbulence,Zhou:PRA1988,Zhou:PR2010}. Thus, we avoid noise renormalization. In this self-consistent approach, we renormalize only the viscosity.  

\begin{figure}[t!]
	\includegraphics[scale=0.7]{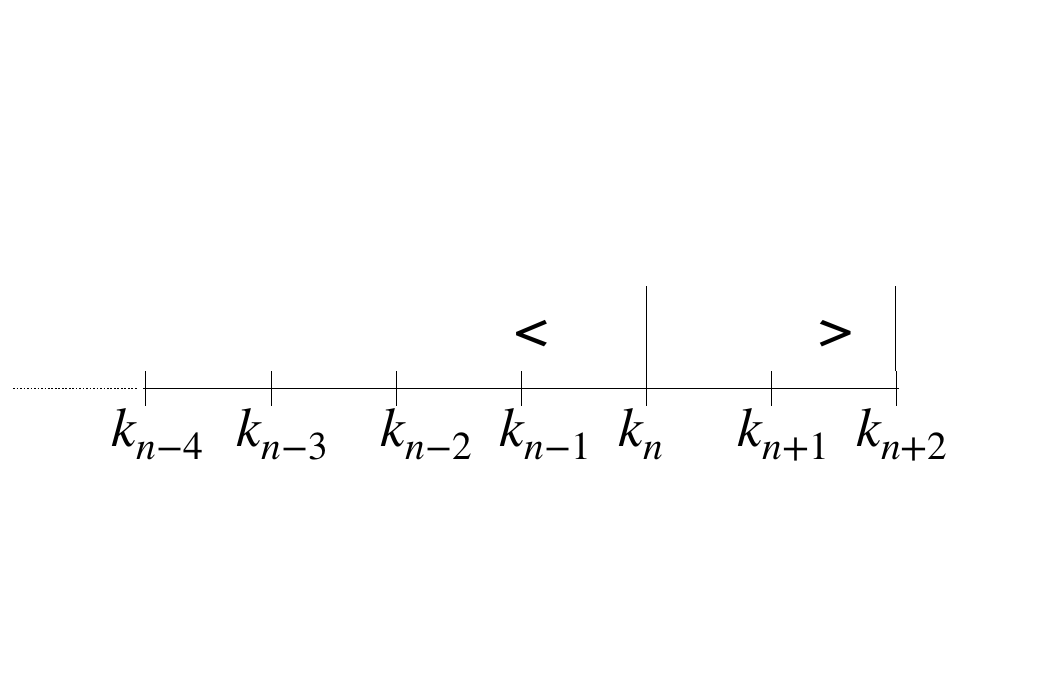}
	\caption{ Division of wavenumber shells into $<$ and $>$ partitions during the computation of $\nu_n$ at $k = k_n$. Under the coarse-graining, the $u^<$  variables are unaltered, whereas $u^>$ variables  are averaged out. } 
	\label{fig:RG_shells}
\end{figure}

Following Wilson~\cite{Wilson:RMP1983}, we coarse-grain the system over a wavenumber shell, and compute the consequent correction to the viscosity.  The wavenumber space is already divided in  the shell model of turbulence, which makes the computation  simpler than that for Navier-Stokes equation.  The locality of interactions too simplifies the RG calculation.  We denote the renormalized viscosity at wavenumber $  k_n  $  by $ \nu_n $.

{  Renormalization is often performed in $ ({\bf k},  \omega)$ space. However, for the shell model, the  renormalization calculation in $ ({\bf k},  t)$ space is concise and convenient. Hence, we adopt this scheme.  In Appendix B, we will briefly discuss the renormalization of the shell model in $ ({\bf k},  \omega)$ space.}

For computing the renormalized viscosity at $k_n$ in the inertial range where $f_n=0$, we coarse-grain the system by averaging over $u_{n+1}^{*>}(t)$ and $ u^>_{n+2}(t)$ (see Fig.~\ref{fig:RG_shells}). Following  RG convention, we label the variables to be averaged using $>$ symbol, whereas those to be retained using $<$ symbol.  Under this notation,
\bea
\left( \frac{d}{dt} + \bar{\nu}  k_n^2 \right) u^<_n(t)  & = & -i   [a_1 k_n u_{n+1}^{*>}(t) u^>_{n+2}(t) \nonumber \\
&& +  a_2 k_{n-1} u_{n-1}^{*>}(t) u^>_{n+1}(t)
\nonumber \\
&&  - a_3 k_{n-2} u^<_{n-1}(t) u^<_{n-2}(t)].
\label{eq:shell_model_nun}
\eea
The  variables with $<$ superscript remain unaltered under coarse-graining.  However, $u^>_{n+1}(t)$ and $ u^>_{n+2} (t)$ variables  are assumed to  be random with zero mean.    Note that $ u^> $ variables need not be Gaussian. Under these assumptions,  
\bea
 \la u^{*<}_{n-1}(t) u^>_{n+1}(t) \ra &=&   u^{*<}_{n-1}(t) \la u^>_{n+1}(t) \ra = 0,  \\
  \la u^<_{n-2}(t) u^<_{n-1}(t) \ra & = & u^<_{n-2}(t) u^<_{n-1}(t)  .
 \eea
 Based on the above simplification,
 \bea
 & \left( \frac{d}{dt} +\bar{\nu} k_n^2 \right) u^<_n(t)  - i a_3 k_{n-2} u^<_{n-1}(t) u^<_{n-2}(t) \nonumber \\
& = -i   a_1 k_n \la u_{n+1}^{*>}(t) u^>_{n+2}(t)  \ra .
  \label{eq:SM_RG_12}
 \eea
 
 To compute the right-hand side (RHS) of Eq.~(\ref{eq:SM_RG_12}), we evaluate $u_{n+1}^{*>}(t)$ and $ u^>_{n+2}(t) $ using Green's function technique. For example,  
 \bea
 u_{n+2}^>(t) & = & \int_{0}^t dt' G_{n+2}(t-t') \times (-i)   [a_1 k_n u_{n+3}^{*>}(t') u^>_{n+4}(t') \nonumber \\
 && +  a_2 k_{n-1} u_{n+1}^{*>}(t') u^>_{n+3}(t')
 \nonumber \\
 &&  - a_3 k_{n} u^<_{n}(t') u^>_{n+1}(t')], 
 \eea
 where $G_{n+2}(t-t')$ is the Green's function. 
 Note, however, that $u_{n+3}^{*>}(t)$ and $ u^>_{n+4}(t)$ are absent at this stage. Hence, 
  \bea
 u_{n+2}^>(t) & = & \int_{0}^t dt'  G_{n+2}(t-t') i a_3 k_{n} u^<_{n}(t') u^>_{n+1}(t'). 
 \label{eq:u_np2}
 \eea
 Substitution of Eq.~(\ref{eq:u_np2}) in the RHS of Eq.~(\ref{eq:SM_RG_12}) yields 
\bea
I_1 & =& \int_{0}^t dt'  G_{n+2}(t-t') a_1  a_3 k_{n}^2 u^<_{n}(t')
 \la u^{*>}_{n+1}(t) u^>_{n+1}(t') \ra  \nonumber \\
& =&  a_1 a_3 k_{n}^2  \int_{0}^t dt'  G_{n+2}(t-t')   
  \bar{C}_{n+1}(t-t') u^<_{n}(t'),
  \label{eq:I1}
\eea
where $\bar{C}_{n+1}(t-t') $ is unequal time correlation.

In self-consistent RG procedure, it is assumed that the decay rates of Green's and correlation functions  are determined by the renormalized viscosity~\cite{Leslie:book,McComb:book:HIT}. Hence, 
 \bea
 G_n (t-t') & = & \theta(t-t')\exp[-\nu_n k_n^2 (t-t')], 
 \label{eq:G_tt'} \\
 \bar{C}_n(t-t') & = & C_n(t) \exp[-\nu_n k_n^2 (t-t')], \label{eq:C_tt'}  
 \eea
 where $C_n(t) $ is equal-time correlation ($t=t'$), and $\theta(t-t')$ is the step function.  Note  that $ G_n (\tau) $ and $ \bar{C}_n (\tau) $ decay with a time scale of $ \tau_c = (\nu_n k_n^2)^{-1}$. Equations~(\ref{eq:G_tt'}, \ref{eq:C_tt'}) are valid for $ \tau < \tau_c $, after which $ C_n$ and  $G_n $ decay rapidly to zero~\cite{Orszag:CP1973,Pope:book,Sanada:PF1992,Verma:INAE2020_sweeping,Zhou:PR2021,Verma:INAE2020_sweeping}. 
 
Substitution of $G_n(t-t')$ and $\bar{C}_n(t-t')$ of Eqs.~(\ref{eq:G_tt'}, \ref{eq:C_tt'}) in Eq.~(\ref{eq:I1}) yields
\bea
I_1 & =&   a_1 a_3 k_{n}^2  C_{n+1}(t) \nonumber \\
&& \times \int_{0}^t dt'   \exp[-(\nu_{n+1} k_{n+1}^2 +\nu_{n+2} k_{n+1}^2) (t-t') ] u^<_{n}(t'),
\label{eq:I1_2} \nonumber \\
\eea
Now, we employ Markovian approximation, according to which the integral of Eq.~(\ref{eq:I1_2}) gets maximal contributions from $t'$ near $t$ \cite{Orszag:CP1973}.  This is possible when $\nu_n k_n^2 \gg 1$~\cite{Orszag:CP1973}.  Since the integral is peaked   near $t=t'$, $u_n(t') \rightarrow u_n(t)$ and 
\bea
I_1 & = & \frac{a_1 a_3 k_n^2 C_{n+1}(t)}{\nu_{n+1} k_{n+1}^2 + \nu_{n+2} k_{n+2}^2} u^<_{n}(t).
\label{eq:I2} 
\eea
Such assumptions are made in Eddy-damped Quasi-normal Markovian (EDQNM) approximation of hydrodynamic turbulence~~\cite{Orszag:CP1973}.

The RHS of Eq.(\ref{eq:SM_RG_12}) has another contribution to $\nu_n$, which is computed   by expanding $u^{*>}_{n+1}(t)$ using Green's function.  Following similar approach as above, we compute new term as
\bea
I_2 & = & \frac{a_1 a_2 k_n^2 C_{n+2}(t)}{\nu_{n+1} k_{n+1}^2 + \nu_{n+2} k_{n+2}^2} u^<_{n}(t).
\label{eq:I1_3} 
\eea	
 
 \begin{figure}[t!]
 	\includegraphics[scale=0.45]{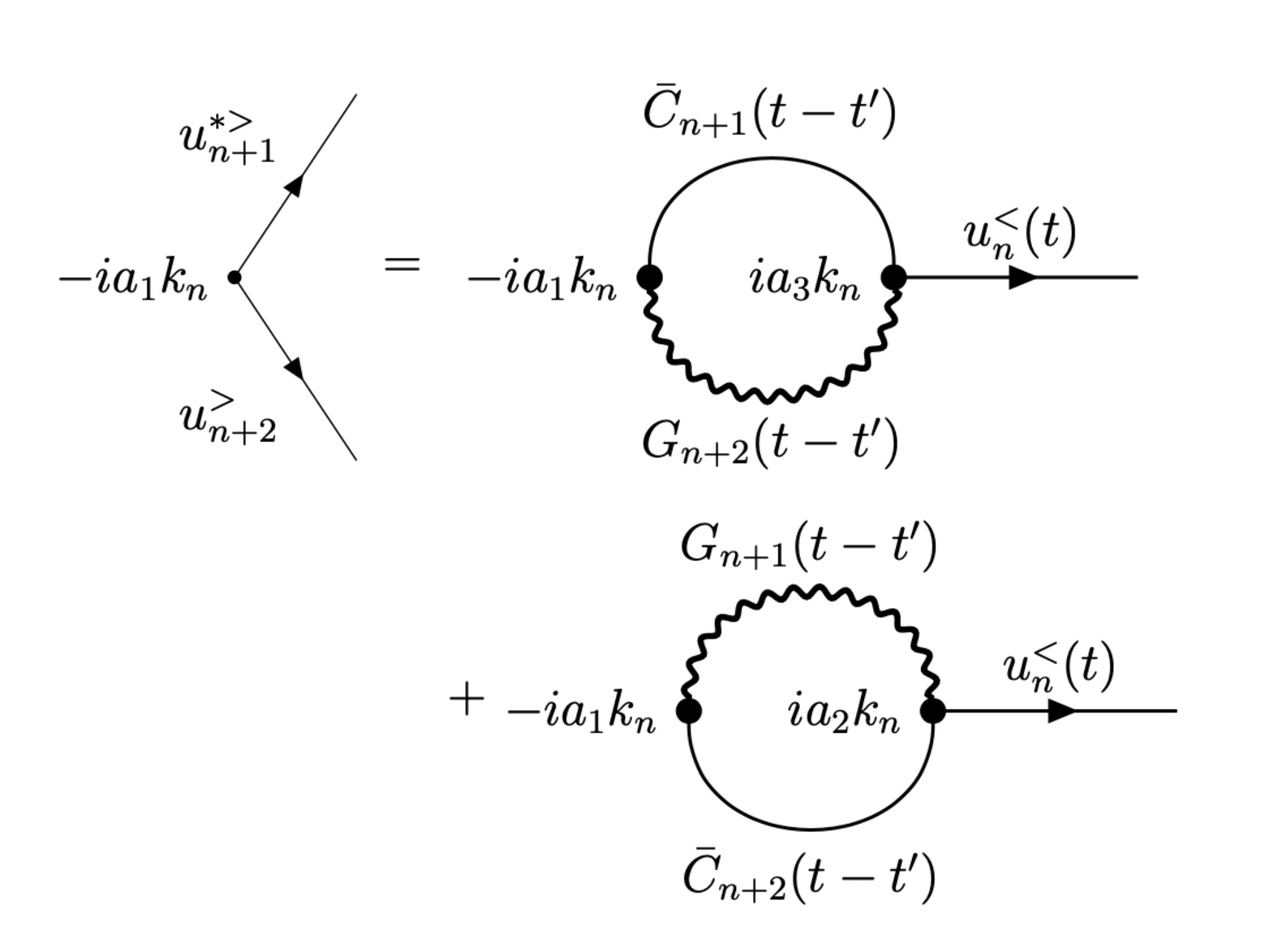}
 	\caption{Feynman diagrams associated with the viscosity renormalization.  These diagrams are related to the RHS of Eq.~(\ref{eq:SM_RG_12}).}
 	\label{fig:feyn_RG}
 \end{figure}
The Feynman diagrams associated with $I_1$ and $I_2$ are exhibited in Fig.~\ref{fig:feyn_RG}.  Here, the loop-diagrams represent  the \textit{self-energy} in which the wavy and solid lines are the Green's function and correlation function respectively. 

These calculations reveal that the  RHS of Eq.(\ref{eq:SM_RG_12}) is proportional to $u^<_n$. Hence, the prefactors of $I_1$ and $I_2$ will provide corrections to $ \bar{\nu} $ to yield $\nu_n$. That is,
\bea
\nu_n k_n^2 =
\bar{\nu} k_n^2 - \frac{a_1 k_n^2 [a_3 C_{n+1}(t) +a_2 C_{n+2}(t)] }{\nu_{n+1} k_{n+1}^2 + \nu_{n+2} k_{n+2}^2}.
\eea
Note, however, that $\bar{\nu}  \ll \nu_n$. Hence,
\bea
\nu_n k_n^2 =  - \frac{a_1 k_n^2 [a_3 C_{n+1}(t) +a_2 C_{n+2}(t)] }{\nu_{n+1} k_{n+1}^2 + \nu_{n+2} k_{n+2}^2}.
\label{eq:RG_eqn}
\eea
Note that we compute renormalized viscosity at the corresponding coarse-graining step. At the present level, $\nu_{n+1}, \nu_{n+2}, ...$ have been computed already, whereas, $\nu_{n-1}, \nu_{n-2}, ...$ would be computed at subsequent stages.  Also note that during the computation of $\nu_{n-1}$, $u^>_{n}$ and $u^>_{n+1}$ would belong to $>$ shells.

In Eq.~(\ref{eq:RG_eqn}), $ \nu_n $ and $ C_n $ are both unknowns. RG equation for Navier-Stokes equation too has a similar implicit form. \citet{Zhou:PRA1988}, and \citet{McComb:PRA1983} employed self-consistent procedure to solve such an implicit equation (also see \cite{McComb:book:Turbulence,Zhou:PF1993,Zhou:PR2010}). Following these authors, we attempt the following functions for $ C_n(t) $ and $ \nu_n $, which are inspired by Kolmogorov's theory of turbulence:
\bea
C_n(t) & = & K_\mathrm{Ko} \epsilon^{2/3} k_n^{-2/3} ,
\label{eq:C_n_formula} \\
\nu_n k_n^2 & = & \nu_* K_\mathrm{Ko}^{1/2} \epsilon^{1/3} k_n^{2/3},
\label{eq:nu_n_formula}
\eea
where $ K_\mathrm{Ko} $ is Kolmogorov's constant, $\epsilon$ is the viscous dissipation rate, and $ \nu_*  $ is the RG constant associated with $ \nu_n $. Substitution of the above in Eq.~(\ref{eq:RG_eqn}) yields
\be
\nu_{*}^2  = - \frac{a_1 (a_3 b^{-2/3} + a_2 b^{-4/3})}{b^{2/3} + b^{4/3}}.
\label{eq:nu_*}
\ee

\begin{figure}[t!]
	\includegraphics[scale=1]{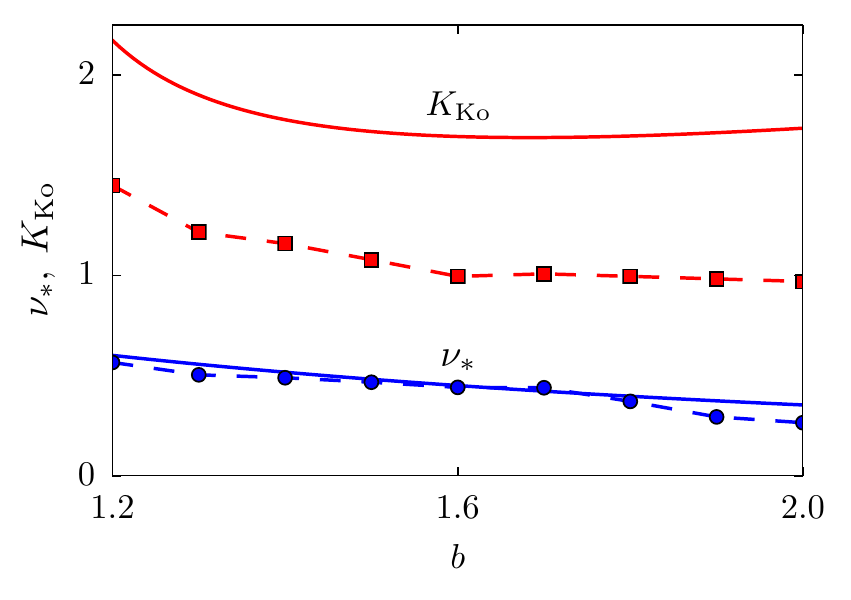}
	\caption{ For the shell model with various $ b $'s, the RG constant $ \nu_* $ computed using RG (solid blue curve) and using numerical simulations (blue circles). Also,  Kolmogorov's constant $ K_\mathrm{Ko} $ computed using field theory (solid red curve) and using numerical simulations (red squares).	The analytical and numerical $ \nu_* $'s match quite well, but numerical $ K_\mathrm{Ko} $ is around 1.6  times smaller than the analytical counterpart.} 
	\label{fig:nu_Ko}
\end{figure}

In Fig.~\ref{fig:nu_Ko} we plot $ \nu_* $ for $ b $ ranging from 1.2 to 2.0. Here, $ \nu_* \approx 0.5 $, in particular, $ \nu_* \approx 0.48 $ for $ b=1.5 $. The $ \nu_* $ computed above are remarkably close to that for Navier-Stokes equation~\cite{McComb:PRA1983,Zhou:PRA1988,	McComb:book:Turbulence,Zhou:PF1993,Zhou:PR2010,Verma:arxiv2005,Verma:PR2004}, which gives credence to the  RG computation described in this paper.

It is important to note that the above derivation does not require quasi-Gaussian assumption for $ u^> $ variables.  We only need to assume time-stationarity for these variables. In addition, we approximate $ \la u^< u^> \ra = u^< \la u^> \ra =0 $, rather than expanding it further.  These assumptions and local interactions in the shell model provide simplification in comparison to the RG calculations for the Navier-Stokes equation~\cite{McComb:PRA1983,Zhou:PRA1988,	McComb:book:Turbulence,Zhou:PF1993,Zhou:PR2010}.

{   Equation (\ref{eq:nu_n_formula}) yields 
\be
 \nu_{n+1} / \nu_n =  (k_{n+1}/k_n)^{-4/3}  = b^{-4/3}.
 \ee
 As is customary in quantum field theory~\cite{Peskin:book:QFT}, we make a change of variable as $b=\exp(l)$, with which
 \be
 \nu_{n+1} = \nu_n \exp(-4l/3) \approx 
 \nu_n [1- 4l/3],
 \ee
 when $  b\rightarrow 1 $ or $l \rightarrow 0$. Hence, 
\be
 \frac{d \nu}{d l} \approx -\frac{4}{3} \nu.
 \label{eq:dnu_dl}
\ee
Therefore, $ \nu_n $ increases with the decrease of $ k_n $, akin to running coupling constant in quantum chromodynamics. Note, however, that $ \nu_n $ is not the coupling constant; instead, it is the coefficient of the viscous term, which is linear (analogous to mass term in quantum field theory). We remark that the scaling of Eqs.~(\ref{eq:C_n_formula}, \ref{eq:nu_n_formula}, \ref{eq:dnu_dl}) breaks down when $ k \rightarrow 1/L $, where $ L $ is the system size.

The dominant frequency at $k=k_n$ is 
\be
\omega_n \sim \nu_n k_n^2 \sim \epsilon^{1/3} k_n^{2/3}.
\ee
For small $k_n$, $\omega_n \rightarrow 0$.  This is one of the assumptions of RG schemes in $({\bf k},\omega)$ space.  Refer to Appendix B for details.}

For $ \bar{\nu}=0 $, $f_n = 0$, and  $ \delta$-correlated (white noise) initial condition, $ u_n $ remains $ \delta$-correlated, as in Euler turbulence~\cite{Kraichnan:PF1964Eulerian,Verma:arxiv2020_equilibrium,Verma:PRF2022}.  Therefore, $ \la u_{n+1}^{*>}(t)  u_{n+2}^>(t) \ra  = 0$ [see Eq.~(\ref{eq:SM_RG_12})],  leading to no correction or renormalization of the viscosity. Thus,  $ \nu_n =0 $ for the inviscid shell model. This solution corresponds to the \textit{Gaussian fixed point} in Wilson's $ \phi^4 $ theory~\cite{Wilson:PR1974}.

In Sec.~III, we will compute the energy flux for the shell model using field-theoretic techniques.

%%%%
\section{Energy flux computation} 
In this section, we  compute the energy flux for the shell model perturbatively. The energy flux at $ k=k_n $ is defined as~\cite{Ditlevsen:book,Biferale:ARFM2003,Verma:book:ET} 
\bea
\Pi_n  & = & 2 a_3 k_{n-1} \Im [\la u^*_{n-1}(t) u^*_n(t) u_{n+1}(t) \ra]
\nonumber  \\
&& -2 a_1 k_{n} \Im [\la u^*_{n}(t) u^*_{n+1}(t) u_{n+2}(t) \ra].
\label{eq:flux}
\eea
We compute $ \la \Pi_n \ra $ by averaging Eq.~(\ref{eq:flux}) under the assumption that $ u_n(t) $'s in the inertial range are quasi-Gaussian with zero mean, an assumption used in eddy-damped quasi-normal Markovian (EDQNM) approximation and in direct interaction approximation (DIA)~\cite{Kraichnan:JFM1959,Orszag:CP1973}.  To zeroth order, $ \la \Pi_n \ra =0 $, which is the energy flux for Euler turbulence; this flux corresponds to the Gaussian  fixed point, $\nu=0$. 

 \begin{figure}[t!]
	\includegraphics[scale=0.37]{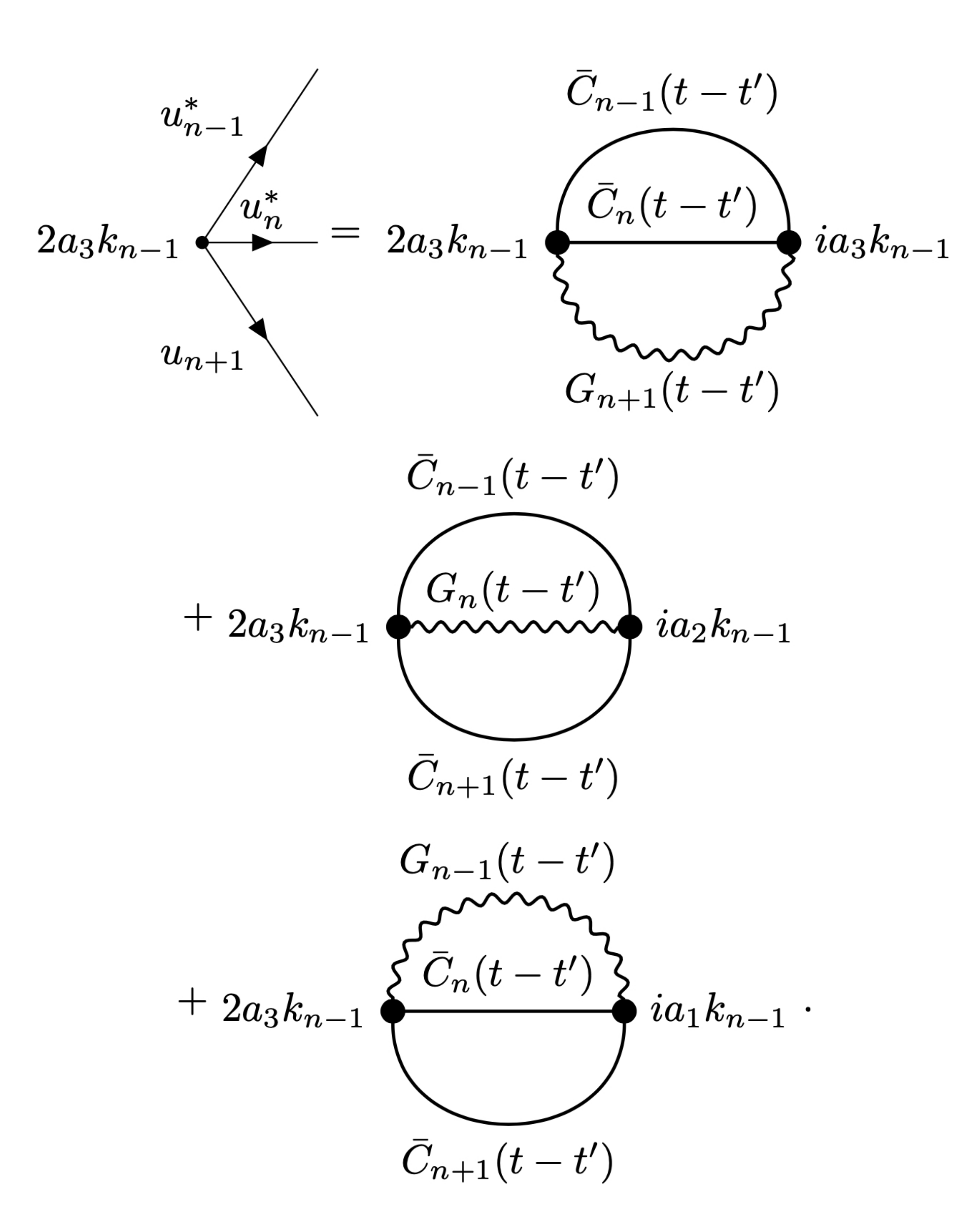}
	\caption{Feynman diagrams associated with the first term of Eq.~(\ref{eq:flux}).} 
	\label{fig:feyn_flux1}
\end{figure}
\begin{figure}[t!]
	\includegraphics[scale=0.37]{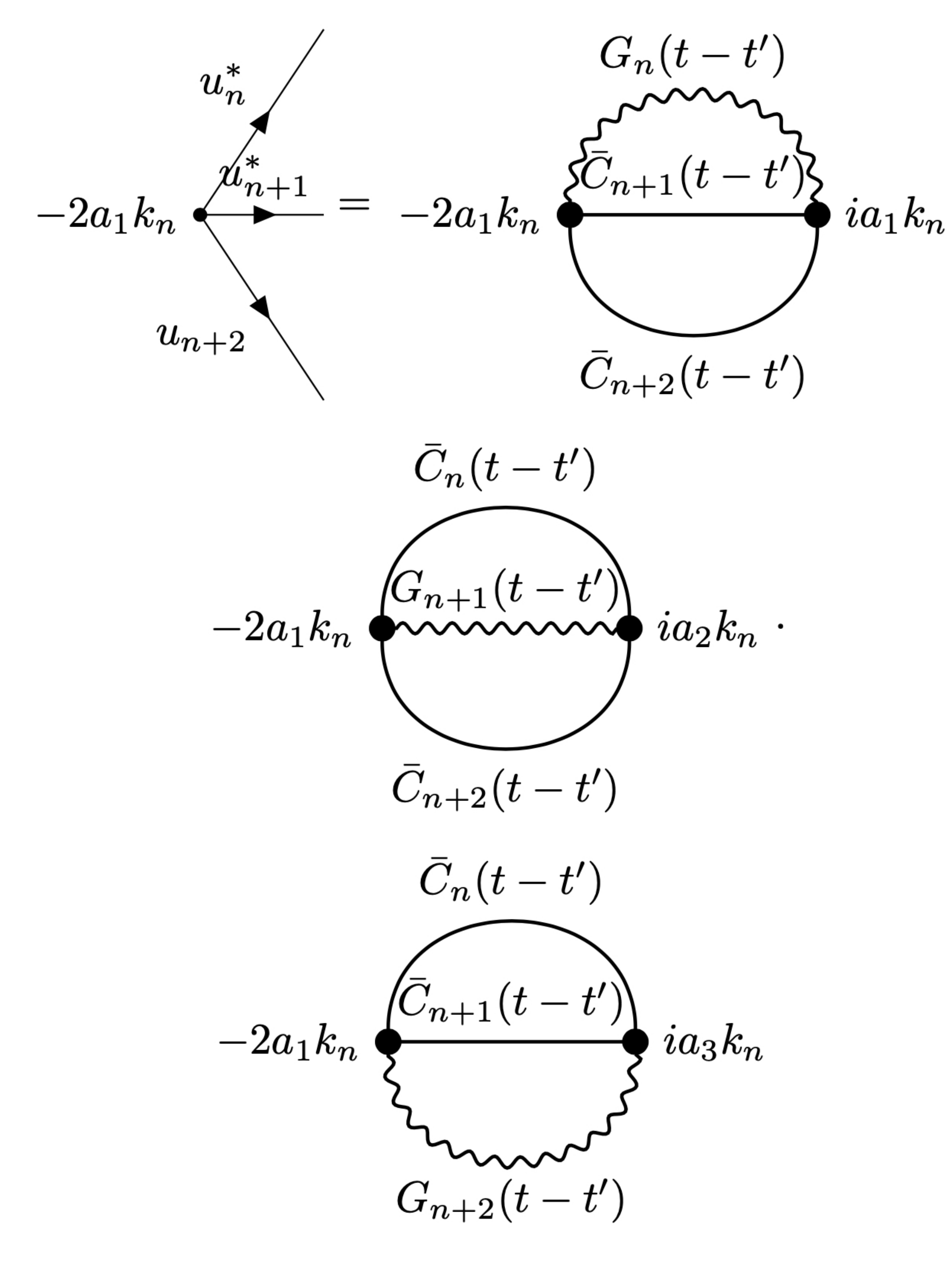}
	\caption{Feynman diagrams associated with the second term of Eq.~(\ref{eq:flux}).} 
	\label{fig:feyn_flux2}
\end{figure}

However, $ \la \Pi_n \ra \ne 0 $ to the first order of perturbation.   The Feynman diagrams associated with the first order in perturbation for the first and second terms of  Eq.~(\ref{eq:flux}) are exhibited in Figs.~\ref{fig:feyn_flux1} and \ref{fig:feyn_flux2} respectively. Let us analyze the expansion of the first Feynman diagram of Fig.~\ref{fig:feyn_flux1}.  Here, $ u_{n+1}(t) $ has been expanded as
\bea
u_{n+1}(t) & =&  \int_{0}^t dt' G_{n+1}(t-t') [-i  a_1 k_{n+1} u_{n+2}^*(t')  u_{n+3}(t') \nonumber \\
&&  \hspace{0.5in} -i a_2 k_{n} u_{n}^*(t') u_{n+2}(t') \nonumber \\
&&\hspace{0.5in} + i a_3 k_{n-1} u_{n}(t') u_{n-1}(t') ] ,
\eea
substitution of which in the first term of  Eq.~(\ref{eq:flux}), or in the first Feynman diagram of Fig.~\ref{fig:feyn_flux1}, yields
\bea
I_3 & = & 2a_3^2 k_{n-1}^2 \int_{0}^t dt' \exp[-\nu_{n+1} k_{n+1}^2(t-t')] \nonumber \\ 
& & \times \Im[i \la u_{n}^*(t) u^*_{n-1}(t) u_{n}(t') u_{n-1}(t') \ra ] \nonumber \\
& = & 2a_3^2 k_{n-1}^2 \int_{0}^t dt' \exp[-\nu_{n+1} k_{n+1}^2(t-t')]  \nonumber \\
&& \times \la u_{n}^*(t)  u_{n}(t') \ra \la  u^*_{n-1}(t)u_{n-1}(t') \ra \nonumber \\
& = & 2a_3^2 k_{n-1}^2  \frac{C_{n}C_{n-1}}{\nu_{n-1} k_{n-1}^2 + \nu_{n} k_{n}^2+\nu_{n+1} k_{n+1}^2 }.
\eea
In the above derivation, we use the following properties:
\begin{enumerate}
\item $ \la u_n(t) u^*_m(t') \ra = \delta_{n,m} C_n(t) \exp[-\nu_{n} k_{n}^2(t-t')] $.
	
\item $ \la a b c d \ra = \la a b \ra  \la c d \ra + \la a c \ra  \la b d \ra + \la a d \ra  \la bc \ra$ when $ a, b, c, d $ are Gaussian variables.
\end{enumerate}
Using similar analysis, we derive the other integrals of the energy flux as
\bea
I_4 & = & 2a_2 a_3 k_{n-1}^2 C_{n-1} C_{n+1} /\mathrm{denr1.}, \\
I_5 & = & 2a_1 a_3 k_{n-1}^2 C_{n} C_{n+1} /\mathrm{denr1.}, \\
I_6 & = & -2a_1^2 k_{n}^2 C_{n+1} C_{n+2} /\mathrm{denr2.}, \\
I_7 & = &- 2a_1 a_2 k_{n}^2 C_{n} C_{n+2} /\mathrm{denr2.}, \\
I_8 & = & -2a_1 a_3 k_{n}^2 C_{n} C_{n+1} /\mathrm{denr2.}, 
\eea
with
\bea
\mathrm{denr1.} = \nu_{n-1} k_{n-1}^2 + \nu_{n} k_{n}^2+\nu_{n+1} k_{n+1}^2,  \\
\mathrm{denr2.} = \nu_{n} k_{n}^2+\nu_{n+1} k_{n+1}^2+\nu_{n+2} k_{n+2}^2.
\eea
The integrals $ I_3, I_4, I_5 $ correspond to the first term of Eq.~(\ref{eq:flux}), whereas $ I_6, I_7, I_8 $ correspond to the second term of Eq.~(\ref{eq:flux}). By adding $ I_3 $ to $ I_8$ and  using $ k_n = k_0 b^n $, we derive
\be
\la \Pi_n \ra = \epsilon = \frac{K_\mathrm{Ko}^{3/2}}{\nu_*} 
\frac{\mathrm{numr}}{1 + b^{2/3} + b^{4/3}},
\label{eq:flux_inertial}
\ee
where
\bea
\mathrm{numr} & = & 2 a_3 b^{-4/3} (a_1 b^{-2/3} + a_2 + a_3 b^{2/3})
\nonumber \\
& & - 2 a_1(a_1 b^{-2} + a_2 b^{-4/3} + a_3 b^{-2/3}).
\eea

Equation~(\ref{eq:flux_inertial}) reveals that the energy flux is independent  of wavenumber, consistent with Kolmogorov's theory of turbulence~\cite{Kolmogorov:DANS1941Dissipation,Kolmogorov:DANS1941Structure,Frisch:book}. Using Eq.~(\ref{eq:flux_inertial}), we compute $K_\mathrm{Ko} $ and plot it in Fig.~\ref{fig:nu_Ko}.  We observe $K_\mathrm{Ko} $ to be a weak function of $ b $. In particular, for $ b=1.5 $,  $K_\mathrm{Ko} \approx 1.71$, which is close to the theoretical, experimental, and numerical values of Kolmogorov's constant~\cite{Frisch:book,McComb:book:Turbulence}.

\section{Sweeping Effect in Shell Model}

{  \citet{Kraichnan:PF1964Eulerian} showed that large-scale flow structures sweep  smaller ones, a phenomenon called \textit{sweeping effect}.  Here, large-scale velocity structures interact with small-scale ones. \citet{Kraichnan:PF1964Eulerian}  observed that the sweeping effect leads to $k^{-3/2}$ energy spectrum, rather than usual $k^{-5/3}$ spectrum.  To overcome this discrepancy,  \citet{Kraichnan:PF1965Lagrangian_history} proposed Lagrangian‐History Closure Approximation for Turbulence. Note that the shell model involves local interactions, thus drastically reduce the sweeping effect. }

To test the sweeping effect in  the field-theoretic calculation of shell model, we introduce a term $ i U_0 k_n u_n $ in the left-hand side of Eq.~(\ref{eq:shell_model}), where $ U_0 $, a constant, represents the mean flow.  Under renormalization, the above term appears as $ i U_n k_n u_n^<$, where $ U_n $ represents the renormalized parameter corresponding to $ U_0 $. With $ U_n $, the RG flow equation [Eq.~(\ref{eq:RG_eqn})] gets transformed to
\begin{gather}
 i U_n k_n + \nu_n k_n^2   = -
\frac{a_1 k_n^2 [a_3 C_{n+1}+a_2 C_{n+2}] }{\mathrm{denr.}}, \label{eq:RG_eqn_sweeping}  \\
\mathrm{denr.} = i[U_{n+1} k_{n+1} + U_{n+2} k_{n+2}] + \nu_{n+1} k_{n+1}^2 + \nu_{n+2} k_{n+2}^2.
\end{gather}
Using dimensional analysis, we argue that 
\be
U_n = U_* \epsilon^{1/3} k_n^{-1/3},
\label{eq:Un_sweeping}
\ee
 substitution of  Eqs. (\ref{eq:C_n_formula},\ref{eq:nu_n_formula}, \ref{eq:Un_sweeping}) in Eq.~(\ref{eq:RG_eqn_sweeping})   yields 
 \be
 (i U_* + \nu_*)^2 = -\frac{a_1  [a_3 b^{-2/3} + a_2 b^{-4/3}] }{b^{2/3} + b^{4/3}}.
 \label{eq:sweeping_eqn}
 \ee
 The only possible solution of Eq.~(\ref{eq:sweeping_eqn}) is
\be 
U_* = 0,~\mathrm{or}~U_n = 0,
\ee
and $ \nu_* $ is given by the same formula as Eq.~(\ref{eq:nu_*}).  Hence, sweeping effect is absent in the RG calculation of the shell model, and $ \nu(k) $ is independent of $ U_0 $. However, in Sec.~V, we show  that the numerical results deviate from the above prediction.

\section{Numerical Verification}

To test the predictions of the  above field-theoretic calculations, we solve Sabra shell model, Eq.~(\ref{eq:shell_model}), numerically.  We employ 40 shells, $ \nu=10^{-6} $, $ U_0=0 $, and  fourth-order Runge-Kutta (RK4) time marching scheme with $ dt = 10^{-5} $. The shell model is forced randomly at shells $ n=0 $ and 1 so as to provide a constant energy supply rate; we choose $ \epsilon=2 $ for all our runs. To test the dependence of $ \nu $ and $ K_\mathrm{Ko} $ on $ b $, we vary $ b$ from 1.2  to 2 in the interval of 0.1. We also perform another simulation with $ U_0 = 0.5 $ and $ b=1.5 $ to test the field-theoretic predictions on the sweeping effect. We carry out the simulations till 2000 eddy turnover times, and report the energy spectra and fluxes after the system has reached a steady state.

\begin{figure}[t!]
	\includegraphics[scale=1]{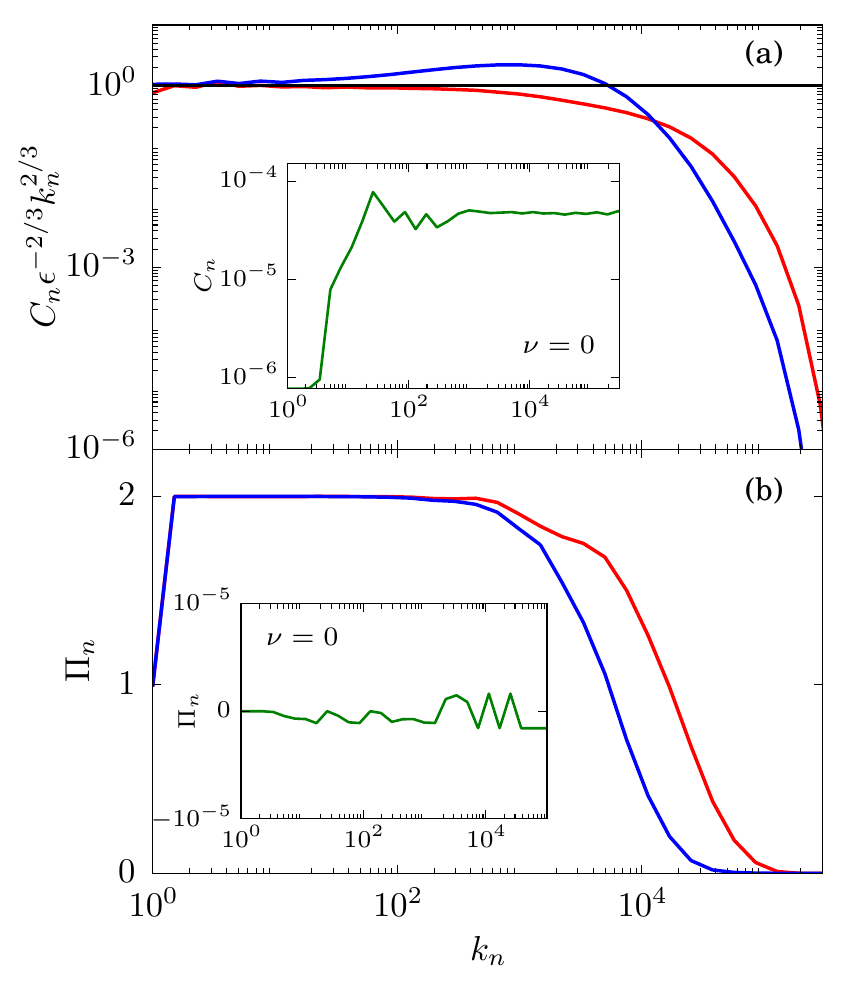}
	\caption{ 	For the numerical simulation of shell model with $ b=1.5 $: (a) plots of normalized energy spectra $ C_n \epsilon^{-2/3} k_n^{2/3} $ vs.~$ k_n $ for $ U_0=0 $ (red curve) and $ U_0=0.5 $ (blue curve). (b) The corresponding energy fluxes $ \Pi_n $ are shown using the same color convention.  We observe $ C_n \sim k_n^{-2/3} $ and constant $ \Pi_n $ in the inertial range. The insets in (a,b) exhibit $ C_n $ and $ \Pi_n $ for the $ \nu=0 $ case (equilibrium behaviour).}
	\label{fig:Ek_Pik}
\end{figure}

As expected, for $ U_0 = 0 $ and finite $ \nu $, in the inertial range, the energy spectrum $ C_n \sim k^{-2/3} $ and the energy flux  $ \Pi_n \approx \epsilon = 2 $ ~\cite{Ditlevsen:book,Biferale:ARFM2003,Plunian:PR2012}. See the red curves of Fig.~\ref{fig:Ek_Pik} for an illustration for $ b=1.5 $.   Using Eq.~(\ref{eq:C_n_formula}), we compute the Kolmogorov's constants for various $ b $'s  and plot them in Fig.~\ref{fig:nu_Ko}. We observe that  for $ b=1.5 $, $ K_\mathrm{Ko} = 1.05 $, which is approximately 1.6 times smaller than the theoretically predicted value of 1.71 (for the shell model). See Fig.~\ref{fig:nu_Ko} for an illustration.    This discrepancy between the numerical and analytical $ K_\mathrm{Ko}  $ is possibly due to various approximations employed in the theoretical calculations, an issue that needs a closer investigation. 

For the   $ U_0=0.5 $ run, we again observe  Kolmogorov's spectrum (apart from a hump) and constant energy flux (blue curves in Fig.~\ref{fig:Ek_Pik}). Here,  $ K_\mathrm{Ko} \approx 0.92 $.   For the special case with $ \bar{\nu}=0 $ and white noise initial condition, numerical simulation yields $ C_n \approx  \mathrm{constant} $ and nearly zero energy flux, consistent with the field-theoretic predictions.  We illustrate the above energy spectrum and flux in the insets of Fig.~\ref{fig:Ek_Pik}.

To validate the renormalized viscosity of Eq.~(\ref{eq:nu_n_formula}), we compute  $ \nu_n $ numerically using the normalized correlation function $ R_n(\tau) $, which is defined as
\be
R_n(\tau) = \frac{\bar{C}(\tau)}{C_n},
\ee
where $\bar{C}(\tau)$ and $C_n$ are the unequal-time and equal-time correlations respectively (see Eq.~(\ref{eq:C_tt'})).

\begin{figure}[t!]
	\includegraphics[scale=1]{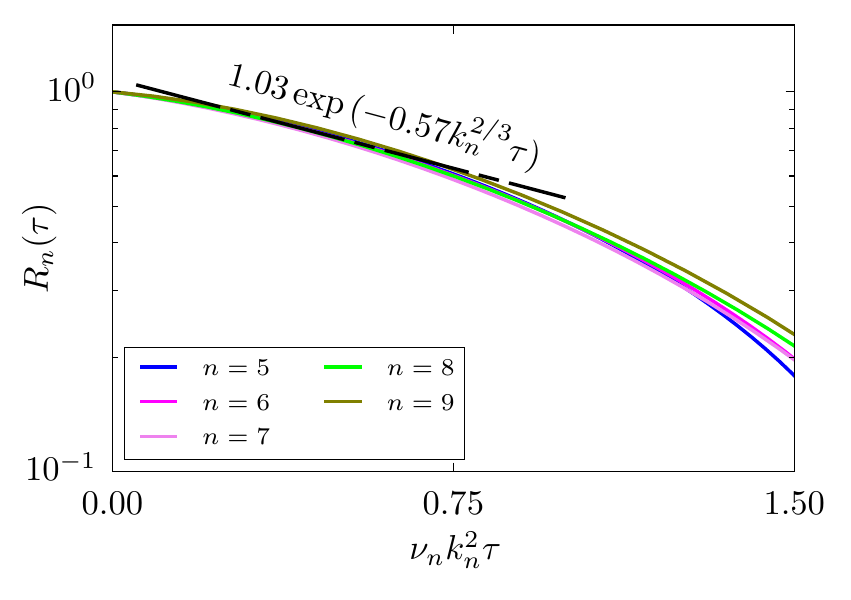}
	\caption{For the shell-model simulation with $ b = 1.5 $ and $ U_0=0 $, plots of $ R_n(\tau) $  vs.~$ \nu_n k_n^{2} \tau$ [Eq.~(\ref{eq:C_tt'})].    The chained straight line represents the best-fit curve $ R_n(\tau) = 1.03 \exp(-0.57 k_n^{2} \tau)$  in the interval $ \nu_n k_n^2 \tau = (0.1,0.75)$.	}
	\label{fig:Ctt'}
\end{figure}

We observe that the numerically-computed $ R_n(\tau) $ is  real.  For small $ \tau $ and inertial range $ k_n $'s, $ R_n(\tau) \approx \exp[-k_n^{2/3} \tau] $, which is consistent with Eqs. (\ref{eq:C_tt'}, \ref{eq:nu_n_formula}). As illustrated in Fig.~\ref{fig:Ctt'},  for $ b=1.5 $ and $ U_0=0 $,
\be 
R_n(\tau) \approx 1.03 \exp(-0.57 k_n^{2/3} \tau)
\label{eq:R_n_numerical}
\ee
when $ \nu_n k_n^2 \tau \lessapprox 1 $. A comparison of Eq.~(\ref{eq:R_n_numerical}) with Eq.~(\ref{eq:C_tt'}) reveals that 
$ \nu_* \approx 0.57/(K_\mathrm{Ko}^{1/2} \epsilon^{1/3}) \approx 0.44$,  which is in good agreement with the RG prediction of 0.48 (see Sec.~II). However, we cautiously remark that the numerical $  \nu_*  $ has significant errors. 

\begin{figure}[t!]
	\includegraphics[scale=1]{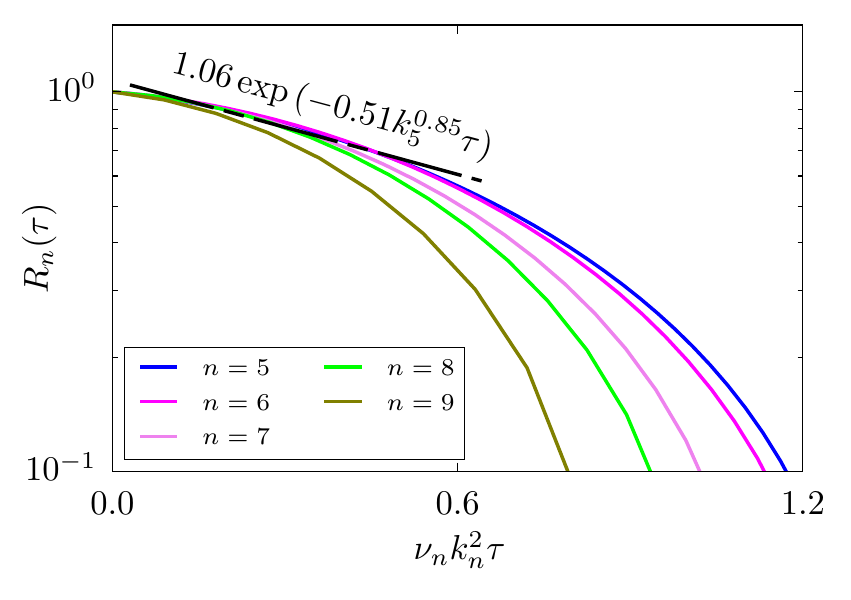}
	\caption{For the shell-model simulation with $ b = 1.5 $ and $ U_0=0.5 $, plots of $ R_n(\tau) $  vs.~$ \nu_n k_n^{2} \tau$.    The chained straight line represents the best-fit curve $ R_n(\tau) = 1.06 \exp(-0.51 k_n^{0.85} \tau)$  in the interval $ \nu_n k_n^{2} \tau = (0.1,0.6)$ for $ n=5 $. Note that the shells  8 and 9 do not follow the best-fit curve.	}
	\label{fig:Ctt'_SM}
\end{figure}

For $ U_0=0.5 $, we compute $ R_n(\tau)$ and fit it with $ \exp(-\nu_n k_n^2 \tau) $.
In Fig.~\ref{fig:Ctt'_SM} we plot $ R_n(\tau) $  for shells $ n=5 $  to 9. We observe that  $ R_n(\tau) $  for $ U_0=0.5 $ is steeper than the RG predictions. This is contrary to the RG prediction that the mean flow does not affect the renormalized viscosity. Clearly, the analytical computation underestimates the dissipation arising due to nonzero $ U_0 $. This issue needs a closer examination that will be pursued in future.

In Sec.~VI, we compare our results with earlier RG works on the  shell model and hydrodynamic turbulence.

\section{Comparison with earlier works}

There are only a small number of works on renormalization group analysis of the shell model. Recently,   \citet{Fontaine:arxiv2022} preformed functional renormalization group (FRG) analysis of the shell model and computed the multiscaling exponents.   They observed that 
\bea
C_n \sim k_n^{-\alpha_E},  \\
\nu_n k_n^2 \sim k_n^{\alpha_\nu},
\eea
with $\alpha_E = 0.633\pm .004$ and $ \alpha_\nu= -0.741 \pm 0.01$. Substitution of the above in renormalization group equation [Eq.~(\ref{eq:RG_eqn})] yields
\be
\alpha_E + 2\alpha_\nu = 2.
\label{eq:coefficient_RG}
\ee
Note that \citet{Fontaine:arxiv2022}'s $\alpha_E $ and $\alpha_\nu $ satisfy Eq.~(\ref{eq:coefficient_RG}) to a good approximation.  \citet{Fontaine:arxiv2022} reported that the proportionality constant for $C_2(\tau=0)$, which is $ K_\mathrm{Ko} $, is approximately 1.15. 

 In a different application of field theory, \citet{Eyink:PRE1993} employed operator product expansion to the shell model and computed various correlations and structure functions.    Note, however, that \citet{Fontaine:arxiv2022} and \citet{Eyink:PRE1993}  do not report the RG  constant $ \nu_*$ in their calculation.   We  remark that the multiscaling exponents are related to  the fluctuations in the energy flux, i.e., for $ \la \Pi_n^2 \ra $~\cite{Sreenivasan:ARFM1991,Das:EPL1994,Verma:arxiv2005}.  The self-consistent calculation presented in this paper may be extendible to the computation of $ \la \Pi_n^2 \ra $. 
 
 It is important to compare  our predictions on $\nu_*$ and $K_\mathrm{Ko}$ with the past works on hydrodynamic turbulence. \citet{Yakhot:JSC1986} observed that $ \nu_* = 0.39 $ and $K_\mathrm{Ko} = 1.617$.  \citet{McComb:JPA1985} computed that   $ \nu_* \approx  0.40 $ and $K_\mathrm{Ko} \approx 1.8$. \citet{Zhou:PRA1988} also reported $ \nu_* \approx  0.40 $. Our field-theoretic computation of the shell model yields $ \nu_* \approx  0.50 $ and $K_\mathrm{Ko} = 1.7$, with minor variations depending on the value of $b$.  Using ERG, \citet{Tomassini:PLB1997} showed that $E(k) \sim k^{-1.666\pm 0.001}$, whereas $K_\mathrm{Ko}$ lies in the range of 1.124 to 1.785 depending on the chosen function. 
 Clearly, the shell model predictions are reasonably close to the earlier works on hydrodynamic turbulence.  

There are  subtle differences between  the RG schemes for the shell model and hydrodynamic turbulence. The RG procedure for the shell model does not involve any integral, and hence is simpler than that for HD turbulence.   In addition, we make fewer assumptions in the RG implementation of the shell model. For example, $u^>$ variables are assumed to be time-stationary, but not necessarily quasi-Gaussian. Note that many past RG works assume that $u^>$ is quasi-Gaussian (see e.g., \cite{McComb:book:HIT}). In addition, the RG computation of the shell model is nearly exact. In Eq.~(\ref{eq:SM_RG_12}), we substitute the expansion of $u^{*>}_{n+1}$ and $u^{>}_{n+2}$ one after the other, and then solve for the $\nu_n$  under Markovian approximation.  Also, note that the local interactions in the shell model  suppresses the \emph{sweeping effect} proposed by \citet{Kraichnan:PF1964Eulerian}.

We conclude in the next section.

\section{Conclusions}

In this paper, we employ  RG analysis to the shell model of turbulence, and show that a combination of Kolmogorov's spectrum  $ C_n = K_\mathrm{Ko} \epsilon^{2/3} k_n^{-2/3} $  and $ \nu_n =\nu_* K_\mathrm{Ko}^{1/2} \epsilon^{1/3}    k_n^{-4/3} $ is a solution of the RG flow equation. Our calculations predict that for $ b=1.5 $, $ \nu_* \approx 0.48 $ and $ K_\mathrm{Ko}  \approx 1.71 $, which are in good agreement with the numerical results, except that  the numerical $ K_\mathrm{Ko} $ is around 1.6 times smaller than the theoretical prediction. Note  that the field-theoretic predictions  for the shell model and the Navier-Stokes equation are close to each other~\cite{McComb:PRA1983,Zhou:PRA1988,McComb:book:Turbulence,Zhou:PRA1988}. 

The computation employed in this paper can be easily generalized to the shell models for scalar and magnetohydrodynamic turbulence.  We also believe that the fluctuations in the energy flux for the shell model could be computed using the method outlined in this paper.

\begin{acknowledgments}
	We thank Soumyadeep Chatterjee in help in drawing the Feynman diagrams of the paper. 
We thank anonymous referees for useful comments and suggestions. This work is supported by the project 6104-1  from the Indo-French Centre for the Promotion of Advanced Research (IFCPAR/CEFIPRA), and the project PHY/DST/2020455 by Department of Science and Technology, India. 
\end{acknowledgments} 

\appendix

\section{Galilean invariance leads to non-renormalizibilty of coupling constant}

It can be easily shown that the coupling constant $ \lambda $ of Navier-Stokes equation (NSE)  remain unchanged on renormalization due to Galilean invariance~\cite{Forster:PRA1977,McComb:book:Turbulence,McComb:book:HIT}.  Here, the derivation is reproduced in brief.  

We write the renormalized Navier-Stokes equation as
\be
\partial_t {\bf u}({\bf x},t)  +\lambda  {\bf u}({\bf x},t) \cdot \nabla {\bf u}({\bf x},t) = -\nabla p({\bf x},t) + \nu  \nabla^2 {\bf u}({\bf x},t), 
\label{eq:NSE}
\ee
where ${\bf u}({\bf x},t),p({\bf x},t)$ are the velocity and pressure fields respectively, $\lambda$ is a measure of the nonlinear interaction, and $\nu$ is the kinematic viscosity. Note that $\lambda=1$ for the original NSE, but it may get renormalized under scaling.

We consider two reference frames: \textit{Lab references frame}, where the fluid has mean velocity ${\bf U}_0 = U_0 \hat{x}$, and the \textit{moving reference frame}, where the velocity is $ {\bf u'}({\bf x'},t') $ with zero mean.  We denote the variables in the lab frames using unprimed variables, but those in the moving frame using primed variables. The variables in the two reference frames are related to each other via Galilean transformation, which is
\be
x = x' + U_0 t';~~~y = y';~~~ z=z';~~~t=t'; 
\ee
\be
\partial_x = \partial_{x'};~~~\partial_y = \partial_{y'};~~~
\partial_z = \partial_{z'};~~~\partial_t = \partial_{t'}-U_0 \partial_x'; 
\ee
\be
{\bf u}({\bf x},t)  =U_0 \hat{x} + {\bf u'}({\bf x'},t');~~~ 
p({\bf x},t)  = p'({\bf x'},t'),
\ee
substitution of which in Eq.~(\ref{eq:NSE}) yields
\begin{eqnarray}
&\partial_{t'} {\bf u'}({\bf x'},t')   + \lambda [U_0 \hat{x} + {\bf u'}({\bf x'},t')] \cdot \nabla' {\bf u'}({\bf x'},t') \nonumber \\
&- U_0 \partial_{x'}{\bf u'}({\bf x'},t')   = -\nabla' p'({\bf x},t) + \nu  \nabla'^2 {\bf u'}({\bf x'},t'). 
\label{eq:NSE'}
\end{eqnarray}
Note that Eq.~(\ref{eq:NSE'}) is  transformed to Eq.~(\ref{eq:NSE}) in primed variables only if 
\be
\lambda = 1.
\ee
Thus, it has been shown that $\lambda $ is unchanged under RG due to Galilean invariance.  For further discussion, refer to \citet{Forster:PRA1977} and \citet{McComb:book:Turbulence,McComb:book:HIT}.

The shell model is written in Fourier space. Hence, it is not possible to extend the above derivation to the shell model. However, using the analogy between the shell model and Navier-Stokes equation,   it is reasonable to assume that $\lambda=1$ for the shell model,  and that $ \lambda $ remains unaltered under RG operation.

\section{ Renormalization of shell model in $ ({\bf k}, \omega) $ space}

In this Appendix, we will briefly discuss renormalization of the shell model in $ ({\bf k},  \omega)$ space.  Note that the shell model is already divided in ${\bf k}$ space. 
The forward and inverse Fourier transforms of $ u_n $ are defined as follows:
\bea
u_n(t) & = & \int_{-\infty}^\infty \frac{d\omega}{2\pi} u_n(\omega) \exp[-i \omega t] , \\
u_n(\omega) & = & \int_{-\infty}^\infty  dt \hspace{0.1cm} u_n(t) \exp[i \omega t]  .
\eea

Fourier transform of Eq. (\ref{eq:shell_model}) yields the following equation for $u^<_n(\omega)$:
\bea
(-i \omega + \bar{\nu} k_n^2) u_n^<(\omega) & = & 
-i   \int \frac{d\omega'}{2\pi} [ a_1 k_n u_{n+1}^{*>}(\omega'-\omega) u^>_{n+2}(\omega') \nonumber \\ 
&& + a_2 k_{n-1} u^{*<}_{n-1}(\omega'-\omega) u^>_{n+1}(\omega') \nonumber \\
&& -a_3 k_{n-2} u^<_{n-2}(\omega-\omega') u^<_{n-1}(\omega') ].
\label{eq:u_n_expansion_f}
\eea
We perform ensemble averaging over $ u_{n+1}^{>} $ and $ u_{n+2}^{>} $ variables. Following the method of Sec.~II,   we arrive at 
\bea
\la u^{*<}_{n-1}(\omega'-\omega) u^>_{n+1}(\omega') \ra &= &   0,  \\
 \la u^<_{n-2}(\omega-\omega') u^<_{n-1}(\omega') \ra  & = & u^<_{n-2}(\omega-\omega') u^<_{n-1}(\omega').
\nonumber \\
\eea
Consequently, only the first term of Eq.~(\ref{eq:u_n_expansion_f}) yields a nonzero correction to the viscosity. 

The renormalized viscosity receives contributions from the two Feynman diagrams of Fig.~\ref{fig:feyn_RG}.  For the first loop diagram, we expand $ u_{n+2}^>(\omega') $ as follows
\bea
&& (-i\omega'+\nu_{n+2} k_{n+2}^2) u_{n+2}^>(\omega') = \nonumber\\
&&-i   \int \frac{d\omega''}{2\pi} [ a_1 k_{n+2} u_{n+3}^{*>}(\omega''-\omega') u^>_{n+4}(\omega'') \nonumber \\ 
&& + a_2 k_{n+1} a_2 u^{*>}_{n+1}(\omega''-\omega') u^>_{n+3}(\omega'') \nonumber \\
&& -a_3 k_{n} u^<_{n}(\omega'-\omega'') u^>_{n+1}(\omega'') ].
\label{eq:u_np2_expansion}
\eea
Note, however, that $ u_{n+3}^{*>}(\omega) $ and  $ u_{n+4}^{*>}(\omega) $ are absent at this stage of expansion. Hence, only the last term of Eq.~(\ref{eq:u_np2_expansion}) survives. Therefore,
\bea
u_{n+2}^>(\omega') = i a_3 k_{n} \int \frac{d\omega''}{2\pi}  \frac{ u^<_{n}(\omega'-\omega'') u^>_{n+1}(\omega'')}{-i\omega'+\nu_{n+2} k_{n+2}^2},
\label{eq:u_np2_expansion_simplified}
\eea
substitution of which in the RHS of  Eq.~(\ref{eq:u_n_expansion_f})  yields
\bea
X & = & -i  a_1 k_n  \int \frac{d\omega'}{2\pi} \la u_{n+1}^{*>}(\omega'-\omega) u^>_{n+2}(\omega') \ra   \nonumber \\
& = &a_1  a_3 k_{n}^2 \int \frac{d\omega'}{2\pi} \frac{d\omega''}{2\pi}  \frac{ \la u^{*>}_{n+1}(\omega'-\omega) u^>_{n+1}(\omega'')  \ra }{-i\omega'+\nu_{n+2} k_{n+2}^2} \nonumber \\
&& \hspace{1in} \times u^<_{n}(\omega'-\omega'') , \nonumber \\
& = &  \left[ a_1  a_3 k_{n}^2 \int \frac{d\omega'}{2\pi}  \frac{ \la |u^{>}_{n+1}(\omega'-\omega)|^2  \ra }{-i\omega'+\nu_{n+2} k_{n+2}^2} \right] u^<_{n}(\omega)  .
\label{eq:self_energy_integral}
\eea
Since $ \nu_n $ is computed for  a long time limit, we set $ \omega \rightarrow 0 $ in the above integral.  Hence, the  square-bracketed term of Eq.~(\ref{eq:self_energy_integral}) is
\bea
I_1 & = &  a_1  a_3 k_{n}^2 \int \frac{d\omega'}{2\pi}  \frac{ \la |u^{>}_{n+1}(\omega')|^2  \ra }{-i\omega'+\nu_{n+2} k_{n+2}^2}  .
\eea

%The ``dressed" Green's function $ G_n(\omega) $ and correlation function $ C_n(\omega) $ are~\cite{McComb:PRA1983,Zhou:PRA1988,McComb:book:Turbulence,McComb:book:HIT,Zhou:PRA1989,Verma:PR2004,Zhou:PR2021}
%\bea
%G_n (\omega) = \frac{1}{-i \omega + \nu_n k_n^2},
%\label{eq:G_n_scaling}  \\
%\la |u_{n}(\omega')|^2 \ra = C_n(\omega) = \frac{C_n}{-i \omega + \nu_n k_n^2}.
%\label{eq:C_n_scaling}
%\eea
%Note that $ C_n(\tau)  = C_n(-\tau) $,  and that $ G_n (\tau) $ and $ C_n (\tau) $ decay with a time scale of $ \tau_c = (\nu_n k_n^2)^{-1}$~\cite{Orszag:CP1973,Pope:book,Sanada:PF1992,Verma:INAE2020_sweeping,Zhou:PR2021,Verma:INAE2020_sweeping}. 

Now, we employ Wiener-Khinchin theorem to simplify the frequency spectrum as
\be
\la |u^{>}_{n+1}(\omega')|^2  \ra =\int_{-\infty}^\infty d\tau C_{n+1}(\tau) \exp(- i \omega' \tau),
\ee
where $ C_{n+1}(\tau) $ is the correlation function defined in Eq.~(\ref{eq:C_tt'}).  With this,
\bea
I_1 & = &  a_1  a_3 k_{n}^2 \int_{-\infty}^\infty d\tau  \bar{C}_{n+1}(\tau)  \int \frac{d\omega'}{2\pi}  \frac{  \exp(-i \omega' \tau) }{-i\omega'+\nu_{n+2} k_{n+2}^2}  . \nonumber \\
\eea
An application of contour integral over the lower part of $ \omega' $ plane  yields
\bea
I_1 & = &  a_1  a_3 k_{n}^2 C_{n+1} \nonumber \\
&&  \times \int_{0}^\infty d\tau 
\exp[-(\nu_{n+1} k_{n+1}^2 + \nu_{n+2} k_{n+2}^2)\tau ] \nonumber \\
& = & \frac{a_1  a_3 k_{n}^2 C_{n+1}}{\nu_{n+1} k_{n+1}^2 + \nu_{n+2} k_{n+2}^2}.
\eea
The second Feynman diagram of Fig.~\ref{fig:feyn_RG}  yields
\bea
I_2 =  \frac{a_1  a_2 k_{n}^2 C_{n+2}}{\nu_{n+1} k_{n+1}^2 + \nu_{n+2} k_{n+2}^2}.
\eea

The steps beyond this point are same as those described in Sec.~II.


%\bibliographystyle{apsrev4-2}
%\bibliography{/Users/mkv/Dropbox/docs-pub/bib/journal,/Users/mkv/Dropbox/docs-pub/bib/book,/Users/mkv/Dropbox/docs-pub/bib/book_edited,/Users/mkv/Dropbox/docs-pub/bib/book_chapter,/Users/mkv/Dropbox/docs-pub/bib/thesis,/Users/mkv/Dropbox/docs-pub/bib/preprint,/Users/mkv/Dropbox/docs-pub/bib/web,/Users/mkv/Dropbox/docs-pub/bib/conf} %\
\end{document}